\documentclass[conference]{IEEEtran}
\IEEEoverridecommandlockouts
\usepackage{cite}
\usepackage{amsmath}
\usepackage{newtxtext,newtxmath}
\usepackage{graphicx}
\usepackage{booktabs}
\usepackage{array,balance}
\usepackage{multirow}
\usepackage{makecell}
\usepackage{xcolor}
\usepackage{tikz}
\usetikzlibrary{arrows.meta,positioning,fit,backgrounds}
\usepackage[hidelinks]{hyperref}
\hypersetup{pdftitle={Beyond QASecClaw: Agent Scaling, Verifier Feedback, and Evidence Reliability in Software Assurance},pdfauthor={S M Abdulla Al Mamun}}

\definecolor{sblue}{HTML}{2a78d6}
\definecolor{sorange}{HTML}{eb6834}
\definecolor{saqua}{HTML}{1baf7a}
\definecolor{sink}{HTML}{52514e}

\begin{document}

\title{Beyond QASecClaw: Agent Scaling,\\Verifier Feedback, and Evidence Reliability\\in Software Assurance}

\author{
\IEEEauthorblockN{
S M Abdulla Al Mamun\textsuperscript{1},
Akif Islam\textsuperscript{2},
Mohd. Ruhul Ameen\textsuperscript{3}
}
\IEEEauthorblockA{
\textsuperscript{1}Department of Computer Science,
The University of Manchester, United Kingdom\\
\textsuperscript{2}Department of Computer Science and Engineering,
University of Rajshahi, Bangladesh\\
\textsuperscript{3}Department of Computer Science and Engineering,
Varendra University, Bangladesh\\[4pt]
\textsuperscript{1}abdulla5akash@gmail.com,
\textsuperscript{2}iamakifislam@gmail.com,
\textsuperscript{3}ruhul@vu.ac.bd
}
}

\maketitle

\begin{abstract}
Can additional specialist language-model contexts improve software assurance when deterministic tool evidence is already available? We study this question through five experiments spanning authored Java warning controls, OWASP Benchmark assessment, Defects4J test generation, SV-COMP assertion verdicts, and deliberately misleading evidence packages. The evaluated systems use Qwen2.5-Coder-7B-Instruct, with a 1.5B sensitivity analysis, and complete 13,638 model requests. On eight related development warnings, increasing the context count from 1 to 50 increases token use 49-fold without improving unsafe clearance. On 160 OWASP cases, model configurations trade clearance against unresolved outcomes; the ten-context arm assesses only 96 cases. On 32 SV-COMP tasks, raw verifier feedback raises mean correct model verdicts from 12.7 to 22.0, but also raises incorrect verdicts from one to seven; a JBMC-plus-JVM validation baseline returns 26 correct verdicts and six unresolved outcomes. On two sets of 120 evidence packages, models reject 70--100\% of applicable counterexamples, while construction-specific scope checks classify every package correctly. No target defect is discovered in three Defects4J bugs, one unsupported by the test adapter. These descriptive results support deterministic evidence admission for the tested settings, while exposing limits of verdict aggregation, feedback interpretation, and the available evaluation coverage.

\end{abstract}


\section{Introduction}
LLM agents now triage static-analysis warnings~\cite{ameen2026qasecclaw,li2025iris,klieber2026adjudicate}, construct verification harnesses~\cite{li2026evident}, translate code into verifiable specifications~\cite{wu2026mags} and reconcile per-function contracts~\cite{lin2026verdict}. Many of these systems divide the work among specialist roles, and general results suggest that sampling and voting over more agents can raise accuracy~\cite{li2024moreagents}. In assurance, however, the costly error is not a missed answer but a \emph{false clearance}: a real defect dismissed as benign. More voices may add confidence without adding evidence, and recent scaling studies report diminishing or negative returns once coordination overhead and correlated errors are counted~\cite{kim2025scalingagents,yang2026agentscaling}.

Building on the warning-assessment question raised by QASecClaw, we examine additional dimensions: context scaling, verifier feedback and evidence applicability. The title refers to this broader evaluation scope; we do not reproduce or claim to outperform QASecClaw. We ask: \emph{under fixed evidence, when do LLM agent contexts add assurance value over deterministic checks that already exist?} We study four research questions:
\begin{itemize}
\item \textbf{RQ1} Does increasing the number of specialist contexts reduce unsafe clearance without inflating abstention or cost?
\item \textbf{RQ2} Does formal-verifier feedback make model verdicts more reliable than the verifier alone?
\item \textbf{RQ3} Can models recognise formal evidence that does not apply to the program and property under assessment?
\item \textbf{RQ4} Can models discover real defects without a supplied warning?
\end{itemize}

Our contributions are: (i) an executable five-experiment study that separates warning clearance, assertion verdicts, evidence applicability and defect discovery; (ii) a descriptive comparison of 13{,}638 completed model requests with deterministic controls, including abstention, cost and incomplete coverage; and (iii) a controlled challenge exposing \emph{applicability confusion}, where a model rejects applicable defect evidence because the program is unsafe. The contribution is an empirical diagnosis of these configurations, not a new general verifier or a test of every multi-agent architecture. Request manifests and assessment outputs were frozen before their corresponding scoring stages; unsuccessful attempts remain documented. The study is exploratory, not preregistered.

\section{Related Work}
\textbf{Agents for assurance.} Table~\ref{tab:related} places this study among the closest systems. QASecClaw filters Semgrep warnings with an LLM agent and reports an 88.6\% reduction in false positives on OWASP Benchmark at a 3.1\% recall cost~\cite{ameen2026qasecclaw}. Evident keeps the final disposition with Frama-C/Eva and uses LLMs only to build harnesses; its LLM-as-judge baselines missed confirmed bugs with every model tested~\cite{li2026evident}. MAGS and VERDICT obtain guarantees through Dafny translation and contract reconciliation~\cite{wu2026mags,lin2026verdict}, and TraceFix verifies the coordination protocol itself~\cite{xia2026tracefix}.

\textbf{Warning adjudication.} Klieber et al.\ report high recall and specificity using repeated consistency checks and reasoning evaluation~\cite{klieber2026adjudicate}. Their stronger reasoning models and different protocols make model capacity and adjudication policy relevant alternative explanations for our results. Prior studies identify leakage in false-alarm classification~\cite{kang2022falsealarms} and weak generalization on realistic vulnerability data~\cite{ding2025primevul}. We therefore report benchmark scope and do not interpret repeated requests as independent evidence.

\textbf{LLMs with verifiers.} Lemur lets LLMs propose invariants that automated reasoners check~\cite{wu2024lemur}; SpecGen generates verifiable specifications for SV-COMP Java programs~\cite{ma2025specgen}; and local open-weight models paired with ESBMC solve far more tasks than symbolic methods alone~\cite{pirzada2026veribmc}. These systems generate artifacts for checking. Our E3 instead asks a model to interpret an unchanged program and raw JBMC feedback~\cite{cordeiro2018jbmc}; no artifact-generation benefit is evaluated.

\begin{table}[t]
\centering
\caption{Closest systems and the boundary they investigate. Prior results are contextual comparisons, not reproduced baselines.}
\label{tab:related}
\setlength{\tabcolsep}{3pt}
\footnotesize
\begin{tabular}{@{}>{\raggedright\arraybackslash}p{0.21\columnwidth}>{\raggedright\arraybackslash}p{0.34\columnwidth}>{\raggedright\arraybackslash}p{0.37\columnwidth}@{}}
\toprule
Work & Principal mechanism & Relation to this study \\
\midrule
QASecClaw~\cite{ameen2026qasecclaw} & LLM filtering of SAST reports & Warning disposition \\
Evident~\cite{li2026evident} & Validated harness; analyser decides & Deterministic admission \\
MAGS~\cite{wu2026mags} & Dafny translation and repair & Specification scope \\
VERDICT~\cite{lin2026verdict} & Assumed contracts and reconciliation & Modular obligations \\
Kim et al.~\cite{kim2025scalingagents} & Controlled agent architectures & Scaling and task fit \\
This study & Fixed-evidence verdict aggregation & Clearance, abstention and applicability \\
\bottomrule
\end{tabular}

\end{table}

\section{Study Design}
\subsection{Assessment pipeline and decision semantics}
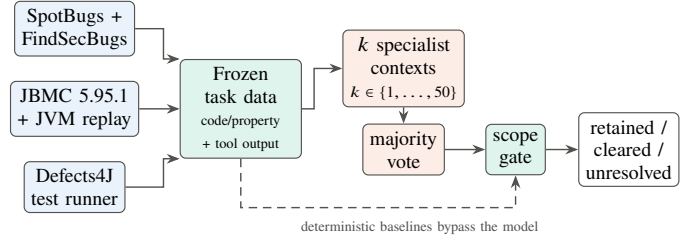
\begin{figure}[t]
\centering
\resizebox{\columnwidth}{!}{%
\begin{tikzpicture}[font=\scriptsize, node distance=3mm and 4mm,
  box/.style={draw=sink, rounded corners=2pt, align=center, inner sep=2.5pt, minimum height=6mm},
  tool/.style={box, fill=sblue!10}, llm/.style={box, fill=sorange!12}, det/.style={box, fill=saqua!14},
  arr/.style={-{Stealth[length=1.6mm]}, draw=sink, line width=0.5pt}]
\node[tool] (scan) {SpotBugs +\\FindSecBugs};
\node[tool, below=of scan] (jbmc) {JBMC 5.95.1\\+ JVM replay};
\node[tool, below=of jbmc] (test) {Defects4J\\test runner};
\node[det, right=5mm of jbmc, minimum width=15mm] (ev) {Frozen\\task data\\{\tiny code/property}\\{\tiny + tool output}};
\node[llm, right=5mm of ev, yshift=5mm] (agents) {$k$ specialist\\contexts\\{\tiny $k\in\{1,\dots,50\}$}};
\node[llm, below=2mm of agents] (vote) {majority\\vote};
\node[det, right=5mm of vote, yshift=-0mm] (gate) {scope\\gate};
\node[box, right=4mm of gate, fill=white] (out) {retained /\\cleared /\\unresolved};
\draw[arr] (scan) -| ([xshift=-2mm]ev.north west) -- (ev.north west);
\draw[arr] (jbmc) -- (ev);
\draw[arr] (test) -| ([xshift=-2mm]ev.south west) -- (ev.south west);
\draw[arr] (ev.east) -- ++(1.5mm,0) |- (agents.west);
\draw[arr] (agents) -- (vote);
\draw[arr] (vote) -- (gate);
\draw[arr] (gate) -- (out);
\draw[arr, densely dashed] (ev.south) |- ([yshift=-4mm]gate.south) -- (gate.south);
\node[font=\tiny, text=sink, align=center] at ([xshift=2mm,yshift=-6.5mm]vote.south) {deterministic baselines bypass the model};
\end{tikzpicture}}
\caption{Overview of the experimental interfaces. Each experiment uses the relevant tool adapter; the tools are not fused into a single end-to-end system. Deterministic baselines bypass the model. E5 outputs accept/reject/unresolved instead of warning dispositions.}
\label{fig:pipeline}
\end{figure}

Fig.~\ref{fig:pipeline} shows the experimental interfaces. Each specialist is a fresh, single-turn context with a role instruction; contexts share a base model and task evidence but do not exchange messages, invoke tools interactively or repair the program. Thus, $k$ counts prompted assessments, not autonomous tool-using agents. The single-context arm uses a general assurance analyst. The five-context arm assigns requirement extraction, assumption identification, data-flow analysis, counterexample interpretation and vacuity critique. Larger teams extend a fixed 50-role catalogue. A strict majority of $k$ returns \textsc{retained} or \textsc{cleared}; otherwise the outcome is \textsc{unresolved}. Missing or invalid answers contribute unresolved votes.

The implemented E1/E4 gate is a restricted postprocessor. Let $S$ indicate matching code and property hashes, and $E$ indicate JBMC success with the declared domain $x\in\{0,\ldots,15\}$ and unwinding assertions enabled. It preserves a proposed clearance only when $S\land E$ holds. Otherwise a matching counterexample changes the clearance to retained; absent such evidence it becomes unresolved. Retentions and abstentions are unchanged. Gate-on and gate-off arms reuse identical model outputs. The controls were checked exhaustively on the JVM before assessment; the gate trusts that preparation and the backend parser. It does not prove harness equivalence or implement the proposed general dependency-graph reconciliation.

E5 uses a different gate for evidence \emph{acceptance}. It checks revision/property hashes, full input-domain coverage, required obligations, exact canonical harness text and whether callee assumptions are marked established. A proposed acceptance with any mismatch becomes rejection. This is a construction-specific validator, not a general proof checker; hashes establish identity, not semantic validity.

\subsection{Metrics}
For labelled defect cases $D$, benign cases $B$, cleared cases $C$, and unresolved cases $U$, we report
\begin{equation}
\mathrm{UC}=\frac{|D\cap C|}{|D|},\quad
\mathrm{BC}=\frac{|B\cap C|}{|B|},\quad
\mathrm{UR}=\frac{|U|}{|D|+|B|}.
\end{equation}
Unsafe clearance (UC) should decrease while benign clearance (BC) increases; unresolved share (UR) exposes abstention. Review workload is the fraction retained or unresolved, not measured reviewer time. Retain-all trivially achieves zero UC but clears no benign cases. We report all three outcomes rather than collapsing them into accuracy. E3 separately scores correct/incorrect/unresolved verdicts; E5 separates corrupted-package acceptance from clean-package rejection; E2 requires a targeted buggy/fixed test differential.

Seeds, variants and packages are repeated or clustered observations. We report counts and seed ranges without significance tests or case-independent confidence intervals. In the ten-context OWASP arm, 64 unassessed cases remain unresolved in the 160-case denominator; we identify these separately from model abstentions. No equal-coverage scaling effect is inferred from that arm.

\subsection{Data, splits and experiments}
Table~\ref{tab:setup} summarises the experiments. Juliet Java 1.3 contains 28{,}881 cases in 112 CWE categories~\cite{nistjuliet13}. We grouped files into 928 family components, merged duplicate complete contexts, and assigned components to development, validation and evaluation (5{,}754 / 2{,}898 / 20{,}229 cases) with no shared testcase source or duplicate complete context across splits. All CWE476 families were reserved for development. This is a dataset audit: no held-out Juliet model result is reported here. The model controls are eight scanner-flagged programs from one authored semantic template, obtained after two earlier constructions yielded no warnings; they are not Juliet testcases. OWASP Benchmark 1.2 (2{,}740 cases, 1{,}415 real vulnerabilities)~\cite{owaspbenchmark} was scanned with SpotBugs 4.8.6 and FindSecBugs 1.14.0~\cite{spotbugs,findsecbugs}; CWE-matched reports covered 1{,}375 of 1{,}415 real cases. For held-out assessment we drew a seeded stratified sample of 160 cases: 60 real and 40 benign cases with a matching report, and 40 real and 20 benign cases without one. E2 used three Defects4J Cli bugs~\cite{just2014defects4j}, each reproduced as a buggy/fixed pair before assessment. E3 used a pinned SV-COMP Java archive~\cite{beyer2025svcomp}: of 1{,}033 manifests, a structural rule fixed before reading verdicts admitted 111 single-file assertion tasks. It requires a default-package \texttt{Main} entry, unused command-line arguments, an assertion property and no explicit imports, threading, native methods or unsupported input APIs; Juliet-derived tasks are excluded. The model comparison takes the first 32 eligible task hashes. These tasks overlap backend calibration, so ``frozen describes the request selection, not an independent backend test set. E5 used 20 null-dereference programs from five templates; each has one \emph{clean} package (a JVM-replayed JBMC counterexample) and five \emph{corrupted} packages whose scoped JBMC success is misleading: restricted precondition, suppressing harness, missing obligation, stale revision and unjustified callee assumption. A second set from five disjoint templates served as held-out replication.

\begin{table}[t]
\centering
\caption{Experiments, data and scale. Req.\ = model requests.}
\label{tab:setup}
\setlength{\tabcolsep}{2pt}
\scriptsize
\begin{tabular}{@{}llrl@{}}
\toprule
Exp. & Data (stage) & Req. & Compared arms \\
\midrule
E1 & 8 authored NPE controls (dev) & 144 & 1 vs 5 ctx; $\pm$gate \\
E4 & same controls, 7B + 1.5B (dev) & 4{,}608 & 1/5/10/25/50; adaptive \\
E1/E4 & 160 OWASP cases (held-out) & 3{,}840 & $k\!\in\!\{1,5,10\}$ \\
E2 & Defects4J Cli-1/2/3 (dev) & 54 & 1 vs 5 ctx; contract tests \\
E3 & 32 SV-COMP tasks (frozen) & 192 & $\pm$feedback; validated backend \\
E5 & 20 progs $\times$ 6 pkgs (dev) & 3{,}600 & $\pm$feedback, $\pm$gate \\
E5 & 20 progs, new templates (held-out) & 1{,}200 & as above \\
\bottomrule
\end{tabular}
\end{table}

\subsection{Models and execution}
All inference used Qwen2.5-Coder-7B-Instruct (revision \texttt{c03e6d3582}) and, for E4 sensitivity, Qwen2.5-Coder-1.5B-Instruct (\texttt{2e1fd397ee})~\cite{hui2024qwen25coder}. The full revision hashes are archived. Sampling used temperature 0.2, top-$p$ 0.95, bfloat16 weights and Transformers 4.51.3. Seeds were 17/43/79 for development and E3, 17/43 for OWASP, and 29 for the second E5 template set. Requests used fresh contexts; batch order and derived batch seeds are recorded, so a seed alone does not ensure cross-hardware reproduction.

For E1/E4 and OWASP, each of $k$ contexts receives at most $\min(8192,\lfloor65536/k\rfloor)$ input tokens and $\min(512,\lfloor8192/k\rfloor)$ output tokens. These are shared upper bounds, not equal actual budgets; the one-context arm has much less aggregate capacity. E3 uses 8,192/512 input/output tokens, E5 8,192/256 per context, and E2 32,768/1,536. E2's five-context arm therefore also has a larger budget. We study the observed cost--outcome trade-off, not a budget-matched causal effect of architecture.

Development ran on university GPUs; the later 5,232 OWASP/E3/E5 requests ran on a Colab A100 in checkpointed batches of 32. Scoring checked frozen request hashes. The 13,638 count is completed final requests, excluding failed attempts and without double-counting gate ablations. E2 additionally records 54 input-cap rejections and 48 out-of-memory failures; the rescue retries failed IDs only. Other failed provisioning attempts are archived separately. A trusted generator renders E2 tests from a restricted declarative schema; arbitrary model-generated Java is not executed.

\section{Results}
\subsection{RQ1: Do more contexts reduce unsafe clearance?}
\textbf{Development controls.} The eight E1 controls comprise four benign and four defective null-pointer warnings, each with matching JBMC evidence. Table~\ref{tab:e1} shows that the ungated model cleared nearly every defect despite being shown the counterexample; inspection confirmed incorrect judgments, not missing context. Five specialist contexts were no better than one. The scope gate removed every unsafe clearance and kept every benign clearance, but the verifier-only baseline already decided all eight correctly.

\begin{table}[t]
\centering
\caption{E1 development controls, Qwen2.5-Coder-7B, three seeds (4 defects and 4 benign warnings per seed).}
\label{tab:e1}
\setlength{\tabcolsep}{4pt}
\footnotesize
\begin{tabular}{@{}lccc@{}}
\toprule
Method & \makecell{Unsafe\\clearance /12} & \makecell{Benign\\clearance /12} & \makecell{Unresolved\\/24} \\
\midrule
Scanner retain-all & 0 & 0 & 0 \\
Verifier only (no model) & 0 & 12 & 0 \\
1 context & 11 & 12 & 1 \\
5 specialist contexts & 11 & 12 & 1 \\
1 context + scope gate & 0 & 12 & 1 \\
5 contexts + scope gate & 0 & 12 & 1 \\
\bottomrule
\end{tabular}
\end{table}

\begin{figure}[t]
\centering
\includegraphics[width=\columnwidth]{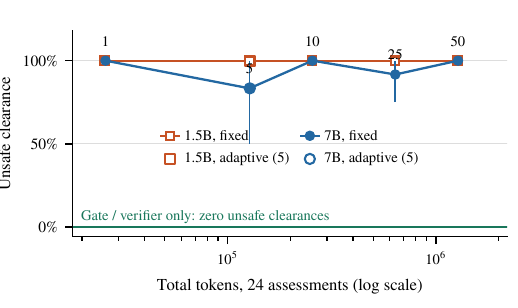}
\caption{E4: unsafe clearance (ungated, mean of 3 seeds; whiskers show the seed range) against total tokens. A 49$\times$ budget increase leaves unsafe clearance near 100\%; the deterministic gate holds it at zero for every $k$.}
\label{fig:scaling}
\end{figure}

\textbf{Scaling.} Fig.~\ref{fig:scaling} plots unsafe clearance against token cost for every E4 configuration. From $k{=}1$ to $k{=}50$ the 7B model's tokens grew from 25{,}938 to 1{,}277{,}612 (49$\times$) while it still cleared 4/4 defects in most seeds; the rare drops came from abstentions, not correct retentions. The 1.5B model cleared every defect in every configuration. The adaptive router activated five roles for every warning because all eight sources matched its trigger, so these data do not test task-dependent routing.

\textbf{Held-out OWASP.} Table~\ref{tab:owasp} gives the joint view on 160 held-out cases. Each context saw the source (capped at 12{,}000 characters), the target CWE and the scanner reports; no verifier evidence exists for this pool, so no gate applies. The clear-without-signal baseline, which clears every case without a CWE-matched report, sits at 40\% unsafe and 33\% benign clearance. A single context reached 33--36\% unsafe and 30--37\% benign clearance, with 45--46\% unresolved. Five contexts raised unsafe clearance to 39--41\%, with 51--53\% unresolved. For ten contexts, 111/113 unresolved outcomes include 64 deliberately unassessed cases per seed; among the 96 assessed cases, 47/49 were unresolved. Its lower full-denominator unsafe clearance is therefore not evidence of a scaling benefit.

At seed 17 the single context improves both clearance rates over clear-without-signal, but raises unresolved share and review workload (105 versus 100 cases). Seed 43 does not improve benign clearance. The data show trade-offs, not equivalence or statistically established superiority. On all 2,740 benchmark cases, clear-without-signal instead yields 2.8\% unsafe and 53.3\% benign clearance: the assessment sample deliberately includes all 40 unflagged real vulnerabilities. Sample rates must not be extrapolated to the full benchmark.

\begin{table}[t]
\centering
\caption{OWASP outcomes on 160 selected cases. Paired counts are seeds 17/43; denominators are fixed across rows.}
\label{tab:owasp}
\setlength{\tabcolsep}{3pt}
\footnotesize
\begin{tabular}{@{}lrrr@{}}
\toprule
Method & \makecell{Unsafe\\/100} & \makecell{Benign\\/60} & \makecell{Unresolved\\/160} \\
\midrule
Retain-all & 0 & 0 & 0 \\
Clear-without-signal & 40 & 20 & 0 \\
1 context & 33/36 & 22/18 & 72/74 \\
5 contexts & 39/41 & 19/23 & 85/81 \\
10 contexts (96 assessed) & 31/28 & 16/19 & 111/113 \\
\bottomrule
\end{tabular}
\end{table}

\textbf{Answer to RQ1.} On the authored controls, added contexts did not improve clearance utility; the gate prevented observed false clearances. OWASP provides no consistent joint improvement, and its ten-context result is confounded by assessment coverage.

\subsection{RQ2: Does verifier feedback help the model?}
We first calibrated JBMC on all 111 eligible SV-COMP tasks (unwind bound 20, 30\,s per call): 64 properties were established under backend models, 25 counterexamples replayed on the JVM, and 22 outcomes were unresolved (12 counterexamples that the JVM did not reproduce, all involving String or library models; 8 different-property failures; one bounded result; one timeout). All 89 decisive outcomes agreed with the expected verdicts.

\begin{figure}[t]
\centering
\includegraphics[width=\columnwidth]{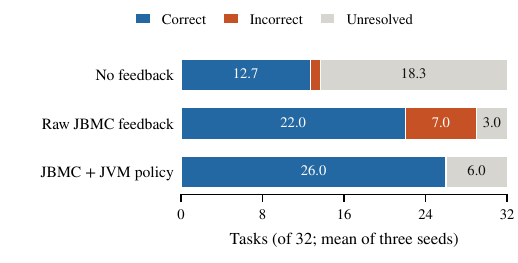}
\caption{E3 verdicts on 32 frozen tasks (means of three seeds). Raw feedback raises both correct and incorrect counts. The deterministic comparator additionally uses JVM validation.}
\label{fig:e3}
\end{figure}

On the 32-task frozen subset (Fig.~\ref{fig:e3}), the model without feedback returned a mean of 12.7 correct, 1 incorrect and 18.3 unresolved verdicts per seed. Feedback increased correct verdicts to 22.0, but also increased incorrect verdicts to 7 in each seed. The deterministic JBMC-plus-JVM policy returned 26 correct and 6 unresolved verdicts. Its aggregate receipt records each task twice, once for each feedback group; we divide those duplicated counts by two rather than reporting 64 independent tasks.

The interfaces differ materially. Model feedback contains the raw JBMC outcome and failed-property records, not the JVM replay outcome or a full validated witness. The deterministic comparator additionally rejects non-replayed assertion failures and failures of other properties. Consequently, this experiment shows that raw feedback did not substitute for that validation policy; it does not establish that the model ignored an identical, fully validated evidence package. It evaluates verdict prediction, not agent-generated invariants or proof strategies.

\subsection{RQ3: Can models recognise misleading evidence?}
\begin{figure}[t]
\centering
\includegraphics[width=\columnwidth]{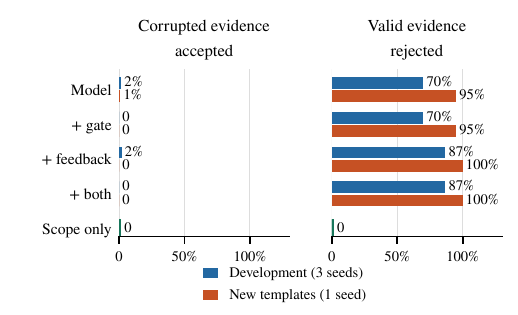}
\caption{E5 misleading-evidence challenge (100 corrupted and 20 clean packages per seed). The gate removes erroneous acceptance but cannot repair the model's rejection of valid evidence; the scope validator alone makes neither error.}
\label{fig:e5}
\end{figure}

Fig.~\ref{fig:e5} separates the two error types. Without the gate the model accepted 0--3\% of corrupted packages, all of them suppressing harnesses or missing obligations; the gate removed these. However, the model rejected 70--100\% of \emph{valid} counterexample packages. Inspected responses rejected a faithful counterexample because the original program still had a null dereference: the model judged the program instead of the evidence. We call this \emph{applicability confusion}. JBMC feedback made it worse (87\% in development, 100\% on held-out templates), and the held-out templates replicated the development pattern. The scope validator alone accepted all 20 clean and rejected all 100 corrupted packages in both sets. Because the corruptions expose their mismatch in metadata or canonical harness text, this baseline solves the known construction; it is not a general semantic-preservation algorithm.

\subsubsection{Worked applicability example}
An actual template uses \texttt{create(x)} to return null when \texttt{x == 0}, and \texttt{entry(x)} dereferences the returned array over inputs 0--15. The clean package analyses that unchanged entry across the entire domain and reports a counterexample. Accepting the package means accepting evidence of a defect, not declaring the program safe. A restricted-domain corruption instead excludes the failing input and obtains solver success; that result is inapplicable to the original full-domain claim. The validator rejects it because the recorded input domain differs, without asking a model to judge the explanation. Other corruptions use a suppressed call, an unrelated obligation, a changed revision or an unestablished callee contract.

This distinction explains why rejecting nearly everything is unhelpful. Without feedback, development clean acceptance was only 3--4/20 per seed; with feedback it was 2--3/20, and on the second template set it fell to 0/20. The gate preserves these rejections because it only vetoes unsupported acceptances. The observed explanatory pattern is therefore a useful diagnostic, although its prevalence in model reasoning was not independently coded.

\subsection{RQ4: Can models discover defects unaided?}
Table~\ref{tab:e2} reports E2. The model saw all 20 production files of each Cli target, with no warning or location, and proposed declarative tests. Under strict parsing, every single-context answer and 41 of 45 five-context answers violated the requested schema; a secondary parser, fixed before any rescue sampling and reported separately, recovered most of them. No generated test, and none of 180 deterministic public-API contract tests, failed on a buggy revision and passed on its fix. Cli-3 concerns typed numeric conversion, which our adapter does not expose, so it remains in the denominator as unsupported. Reproduction on held-out Lang bugs failed under JDK~11 (fixed revisions still fail 8 tests), while six Csv, Codec and Compress bugs reproduced and await assessment.

\begin{table}[t]
\centering
\caption{E2 independent discovery on Defects4J Cli-1/2/3 (3 seeds). Counts are executed test specifications, not independent bugs.}
\label{tab:e2}
\setlength{\tabcolsep}{3pt}
\footnotesize
\begin{tabular}{@{}lrrrr@{}}
\toprule
Setting & Executed & \makecell{Pass\\both} & \makecell{Fail\\both} & \makecell{Buggy/fixed\\differential} \\
\midrule
1 context, strict parser & 0 & 0 & 0 & 0 \\
5 contexts, strict parser & 16 & 13 & 3 & 0 \\
1 context, secondary parser & 28 & 25 & 3 & 0 \\
5 contexts, secondary parser & 125 & 104 & 21 & 0 \\
Contract-test baseline (no model) & 180 & 171 & 9 & 0 \\
\bottomrule
\end{tabular}
\end{table}

\section{Discussion}
\textbf{What the gate demonstrates.} Across the E1 and E4 configurations, postprocessing removed 161 unsafe clearances; across both E5 sets it removed 12 erroneous acceptances. These are repeated decisions, not distinct defects. It removed no correct clearance or package acceptance in these constructions, but inherited model abstentions and clean-package rejections. The stronger tools-only baselines already had the necessary evidence, so the gate's success is a consistency check rather than an independent gain in verification capability. This supports the evidence-controlled disposition used by Evident~\cite{li2026evident} within the boundaries tested here.

\textbf{What scaling demonstrates.} Homogeneous models with repeated evidence can share errors, consistent with prior diversity-based analyses~\cite{yang2026agentscaling}. We did not estimate error correlation or isolate role diversity from sampling. Larger teams also include roles poorly matched to null dereferences and receive different per-context caps. Thus the experiment challenges this fixed catalogue and aggregation rule, not the usefulness of specialised teams with complementary tools, iterative communication or stronger models. A useful next comparison would give a single model the same total inference budget and compare identical evaluated cases.

\textbf{Applicability as a separate outcome.} Evidence can be applicable precisely because it demonstrates failure. Evaluations should therefore measure acceptance of valid counterexamples as well as rejection of invalid assurance claims. E5 exposes that distinction, while its known constructions make deterministic scope checks unusually strong. Tests using independently authored harnesses, applicable proofs of benign programs and semantically equivalent alternative harnesses are needed to measure general applicability reasoning.

\section{Threats to Validity}
\textit{Construct validity.} OWASP is scored at case/CWE level, not by independently adjudicated warning location, and includes unflagged cases; it is a case-assessment experiment rather than pure warning filtering. E5 clean packages are all counterexamples and corrupted packages all solver successes, confounding applicability with outcome type. The canonical-text validator may reject an equivalent unfamiliar harness. E2's adapter cannot exercise Cli-3's target API, and its test counts do not measure semantic diversity.

\textit{Internal validity.} Development controls were revised after zero-warning attempts. E3 compares raw feedback with a baseline that also uses JVM validation. Input/output caps vary with team size, actual token use differs, and OWASP's ten-context arm omits 64 cases. E2's secondary parser was chosen after six completed outputs, before rescue sampling; it is a post hoc diagnostic, not the original primary analysis. Frozen request hashes prevent later input changes but are not preregistration. An independent implementation audit remains outstanding.

\textit{External validity.} Only two small models from one family were evaluated; the second model appears only in the eight-control scaling study. OWASP is one synthetic application, E2 covers one project, and E3 admits 111 of 1,033 manifests. Developer-authored E5 templates are not independently designed challenges. Public benchmark contamination remains possible; source identities were not comprehensively masked. The prepared family-disjoint Juliet split does not establish generalization without model evaluation on it.

\textit{Conclusion validity.} Seeds, templates and packages are dependent, and no effect-size threshold or power analysis was fixed for this exploratory report. We make no significance, equivalence or universal soundness claim. JVM replay independently supports failures; successful JBMC outcomes remain conditional on its library models and parser, with no independently checked proof certificate. Failed attempts and unassessed tasks are distinguished from observed negative results.

\section{Conclusion}
In the tested Java settings, additional single-turn specialist contexts did not produce a consistent assurance benefit over the available deterministic controls. Raw verifier feedback increased correct model verdicts but also increased incorrect verdicts. Construction-specific scope checks prevented observed unsupported acceptances, while model rejection of applicable counterexamples limited utility. These findings favour retaining deterministic evidence admission when the required checks are available. They motivate broader, budget-matched evaluation rather than a general claim against multi-agent software assurance.

\section*{Data Availability}
The private research repository archives executable adapters, model/configuration pins, aggregate receipts, selection manifests and SHA-256 provenance. Figures and tables in this paper derive from those receipts. Raw model responses, benchmark sources and solver traces are retained separately; this manuscript does not claim a public artifact release. A review package requires access arrangements and upstream licence checks. The pending 2,032-case Juliet comparison, six reproduced Csv/Codec/Compress targets and independently authored E5 challenge are not reported as completed experiments.

\section*{Acknowledgment}
Experiments used the University of Manchester CSF3 cluster and Google Colab.

\balance
\bibliographystyle{IEEEtran}
\bibliography{refs}

@inproceedings{kang2022falsealarms,
  author    = {Hong Jin Kang and Khai Loong Aw and David Lo},
  title     = {Detecting False Alarms from Automatic Static Analysis Tools: How Far are We?},
  booktitle = {Proceedings of the 44th International Conference on Software Engineering (ICSE)},
  pages     = {698--709},
  year      = {2022},
  doi       = {10.1145/3510003.3510214}
}

@misc{owaspbenchmark,
  author       = {{OWASP Foundation}},
  title        = {{OWASP} Benchmark Project, version 1.2},
  howpublished = {\url{https://owasp.org/www-project-benchmark/}},
  note         = {Java test suite: \url{https://github.com/OWASP-Benchmark/BenchmarkJava}. Accessed: 2026-10-26}
}

@inproceedings{just2014defects4j,
  author    = {Ren{\'e} Just and Darioush Jalali and Michael D. Ernst},
  title     = {{Defects4J}: A Database of Existing Faults to Enable Controlled Testing Studies for {Java} Programs},
  booktitle = {Proceedings of the 2014 International Symposium on Software Testing and Analysis (ISSTA)},
  pages     = {437--440},
  year      = {2014},
  doi       = {10.1145/2610384.2628055}
}

@misc{spotbugs,
  author       = {{SpotBugs Team}},
  title        = {{SpotBugs}: Find Bugs in {Java} Programs},
  howpublished = {\url{https://spotbugs.github.io/}},
  note         = {Accessed: 2026-10-26}
}

@misc{findsecbugs,
  author       = {{Find Security Bugs Project}},
  title        = {Find Security Bugs: The {SpotBugs} Plugin for Security Audits of {Java} Web Applications},
  howpublished = {\url{https://find-sec-bugs.github.io/}},
  note         = {Version 1.14.0. Accessed: 2026-10-26}
}

@inproceedings{cordeiro2018jbmc,
  author    = {Lucas Cordeiro and Pascal Kesseli and Daniel Kroening and Peter Schrammel and Marek Trtik},
  title     = {{JBMC}: A Bounded Model Checking Tool for Verifying {Java} Bytecode},
  booktitle = {Computer Aided Verification (CAV 2018)},
  editor    = {Hana Chockler and Georg Weissenbacher},
  series    = {Lecture Notes in Computer Science},
  publisher = {Springer},
  pages     = {183--190},
  year      = {2018},
  doi       = {10.1007/978-3-319-96145-3_10}
}

@misc{ameen2026qasecclaw,
  title         = {{QASecClaw}: A Multi-Agent {LLM} Approach for False Positive Reduction in Static Application Security Testing},
  author        = {Ameen, Mohd Ruhul and Alam, Md Takrim Ul and Islam, Akif},
  year          = {2026},
  eprint        = {2605.01885},
  archivePrefix = {arXiv},
  note          = {arXiv preprint arXiv:2605.01885}
}

@misc{xia2026tracefix,
  title         = {{TraceFix}: Repairing Agent Coordination Protocols with {TLA+} Counterexamples},
  author        = {Xia, Shuren and Li, Qiwei and Ehsan, Taqiya and Ortiz, Jorge},
  year          = {2026},
  eprint        = {2605.07935},
  archivePrefix = {arXiv},
  note          = {arXiv preprint arXiv:2605.07935}
}

@misc{li2026evident,
  title         = {The Hitchhiker's Guide to Program Analysis, Part {III}: Mostly Harmless {LLMs}},
  author        = {Li, Haonan and Zhou, Tianyang and Sridharan, Manu and Zhang, Hang and Qian, Zhiyun},
  year          = {2026},
  eprint        = {2606.15122},
  archivePrefix = {arXiv},
  note          = {arXiv preprint arXiv:2606.15122v2}
}

@misc{wu2026mags,
  title         = {{MAGS}: Multi-agent Auto-formalization Guarantees Safety for Agentic Outputs},
  author        = {Wu, Albert and Roberts, Nicholas and Huang, Tzu-Heng and Lin, Haoran and Friedman, Gil and Cho, Sungjun and Orlanski, Gabriel and Sala, Frederic},
  year          = {2026},
  eprint        = {2609.19391},
  archivePrefix = {arXiv},
  note          = {arXiv preprint arXiv:2609.19391}
}

@misc{lin2026verdict,
  title         = {{VERDICT}: Multi-Agent Parallel Modular Verification},
  author        = {Lin, Tengjie and Wang, Zhongyi and Li, Haokun and Yin, Jianwei and Chen, Mingshuai},
  year          = {2026},
  eprint        = {2610.29147},
  archivePrefix = {arXiv},
  note          = {arXiv:2610.29147. Accepted at ICSE 2027}
}

@misc{yang2026agentscaling,
  title         = {Understanding Agent Scaling in {LLM}-Based Multi-Agent Systems via Diversity},
  author        = {Yang, Yingxuan and Qu, Chengrui and Wen, Muning and Shi, Laixi and Wen, Ying and Zhang, Weinan and Wierman, Adam and Gu, Shangding},
  year          = {2026},
  eprint        = {2602.03794},
  archivePrefix = {arXiv},
  note          = {arXiv preprint arXiv:2602.03794}
}

@misc{kim2025scalingagents,
  title         = {Towards a Science of Scaling Agent Systems},
  author        = {Kim, Yubin and Gu, Ken and Park, Chanwoo and Park, Chunjong and Schmidgall, Samuel and Heydari, A. Ali and Yan, Yao and Zhang, Zhihan and Zhuang, Yuchen and Liu, Yun and Malhotra, Mark and Liang, Paul Pu and Park, Hae Won and Yang, Yuzhe and Xu, Xuhai and Du, Yilun and Patel, Shwetak and Althoff, Tim and McDuff, Daniel and Liu, Xin},
  year          = {2025},
  eprint        = {2512.08296},
  archivePrefix = {arXiv},
  note          = {arXiv preprint arXiv:2512.08296}
}

@article{li2024moreagents,
  title         = {More Agents Is All You Need},
  author        = {Li, Junyou and Zhang, Qin and Yu, Yangbin and Fu, Qiang and Ye, Deheng},
  journal       = {Transactions on Machine Learning Research},
  year          = {2024},
  eprint        = {2402.05120},
  archivePrefix = {arXiv}
}

@misc{klieber2026adjudicate,
  author        = {William Klieber and David Svoboda and Lori Flynn and Ruben Martins},
  title         = {Using {LLMs} to Adjudicate Static-Analysis Alerts with Error Reduction Techniques},
  year          = {2026},
  eprint        = {2607.09979},
  archivePrefix = {arXiv},
  primaryClass  = {cs.SE},
  howpublished  = {arXiv preprint arXiv:2607.09979}
}

@inproceedings{li2025iris,
  author    = {Ziyang Li and Saikat Dutta and Mayur Naik},
  title     = {{IRIS}: {LLM}-Assisted Static Analysis for Detecting Security Vulnerabilities},
  booktitle = {Proceedings of the International Conference on Learning Representations (ICLR)},
  year      = {2025},
  note      = {arXiv:2405.17238}
}

@inproceedings{ding2025primevul,
  author    = {Yangruibo Ding and Yanjun Fu and Omniyyah Ibrahim and Chawin Sitawarin and Xinyun Chen and Basel Alomair and David Wagner and Baishakhi Ray and Yizheng Chen},
  title     = {Vulnerability Detection with Code Language Models: How Far Are We?},
  booktitle = {Proceedings of the IEEE/ACM 47th International Conference on Software Engineering (ICSE)},
  pages     = {1729--1741},
  year      = {2025},
  doi       = {10.1109/ICSE55347.2025.00038}
}

@misc{hui2024qwen25coder,
  author        = {Binyuan Hui and Jian Yang and Zeyu Cui and Jiaxi Yang and Dayiheng Liu and Lei Zhang and Tianyu Liu and Jiajun Zhang and Bowen Yu and Keming Lu and Kai Dang and Yang Fan and Yichang Zhang and An Yang and Rui Men and Fei Huang and Bo Zheng and Yibo Miao and Shanghaoran Quan and Yunlong Feng and Xingzhang Ren and Xuancheng Ren and Jingren Zhou and Junyang Lin},
  title         = {{Qwen2.5-Coder} Technical Report},
  year          = {2024},
  eprint        = {2409.12186},
  archivePrefix = {arXiv},
  primaryClass  = {cs.CL},
  howpublished  = {arXiv preprint arXiv:2409.12186}
}

@inproceedings{beyer2025svcomp,
  author    = {Dirk Beyer and Jan Strej{\v{c}}ek},
  title     = {Improvements in Software Verification and Witness Validation: {SV-COMP} 2025},
  booktitle = {Tools and Algorithms for the Construction and Analysis of Systems (TACAS 2025)},
  editor    = {Arie Gurfinkel and Marijn Heule},
  series    = {Lecture Notes in Computer Science},
  volume    = {15698},
  publisher = {Springer},
  pages     = {151--186},
  year      = {2025},
  doi       = {10.1007/978-3-031-90660-2_9}
}

@inproceedings{wu2024lemur,
  author    = {Haoze Wu and Clark Barrett and Nina Narodytska},
  title     = {Lemur: Integrating Large Language Models in Automated Program Verification},
  booktitle = {The Twelfth International Conference on Learning Representations (ICLR 2024)},
  year      = {2024},
  url       = {https://openreview.net/forum?id=Q3YaCghZNt},
  note      = {arXiv:2310.04870}
}

@inproceedings{ma2025specgen,
  author    = {Lezhi Ma and Shangqing Liu and Yi Li and Xiaofei Xie and Lei Bu},
  title     = {{SpecGen}: Automated Generation of Formal Program Specifications via Large Language Models},
  booktitle = {Proceedings of the IEEE/ACM 47th International Conference on Software Engineering (ICSE 2025)},
  publisher = {IEEE},
  pages     = {16--28},
  year      = {2025},
  doi       = {10.1109/ICSE55347.2025.00129},
  note      = {arXiv:2401.08807}
}

@misc{pirzada2026veribmc,
  author        = {Muhammad A. A. Pirzada and Julian Parsert and Weiqi Wang and Konstantin Korovin and Lucas C. Cordeiro},
  title         = {Neuro-Symbolic Software Verification: Hyper-charging Local Language Models with Symbolic Reasoning at Scale},
  year          = {2026},
  eprint        = {2606.16886},
  archivePrefix = {arXiv},
  primaryClass  = {cs.SE},
  doi           = {10.48550/arXiv.2606.16886},
  note          = {arXiv preprint arXiv:2606.16886}
}

@misc{nistjuliet13,
  author = {{National Institute of Standards and Technology}},
  title = {{Juliet Java 1.3}, Software Assurance Reference Dataset, Test Suite 111},
  year = {2017},
  howpublished = {\url{https://samate.nist.gov/SARD/test-suites/111}},
  note = {Accessed: 2026-10-26}
}

\end{document}